\documentclass[12pt]{article}
\usepackage[a4paper,left=28mm,right=25mm,top=30mm,bottom=30mm]{geometry}
\newcommand{\nn}[0]{\hspace*{.7em}}
\newcommand{\n}[0]{\hspace*{.35em}}
\usepackage{natbib}

\begin{document}

\noindent{Observational Studies, 1, (2015), to appear}\\[2mm]

\noindent{{\begin{center}{\large \bf Design and interpretation of studies: relevant concepts from the past and some extensions}\end{center}}
\noindent {\begin{center}{\bf{D.R. Cox* and Nanny Wermuth**}} \end{center}

\noindent \n {\em* Nuffield College, Oxford University, Oxford, United Kingdom\\}
{\em  **Mathematical Statisics, Chalmers University of Technology, Gothenburg, Sweden, and 
\nn \n Medical Psychology and Medical Sociology, Gutenberg-University, Mainz, Germany}\\

We are happy to have the chance of discussing the paper by W.G. (Bill) Cochran, titled `Observational studies' and reprinted here. It appeared first in 1972 and, we call it the `present paper', below. We start however by describing our personal encounters with Bill.

\subsubsection*{1 Personal Encounters with Bill Cochran}
\noindent [DRC] I first heard Bill Cochran lecture in 1956 and, about that time, greatly benefited from his pre-publication comments on a draft of a book on experimental design.  I recall also a memorable meeting of the Royal Statistical Society at which the precursor (Cochran, 1965) of the present paper was given for discussion.\\[-2mm]

\noindent[NW] As a Ph.D. student, I was fortunate to get to know Bill Cochran as an excellent teacher and researcher. His way of teaching was typically most illuminating for me. He was involved in many different types of empirical studies and he shared his experiences openly with the students. He would talk with joy about successes but would also report on disappointing developments that had led to difficult, unsolved problems.
I regarded him as the heart of our department. He stressed the positive features of his colleagues and he remembered the names of all the students as well as what he had discussed with them before. This could concern statistical questions or personal experiences. He was kind and modest, typically full of energy, and always ready to listen and talk. I learned a lot from him not only about statistics.

\subsubsection*{2 Discussion of Cochran's `Observational Studies'}
The present paper is striking for its relevance even after so many years. Cochran's concepts and ideas are presented with clarity and simplicity. Many of them appear to be ignored in the current inrush of `big data'. This makes many of Cochran's points ever more topical.\\[-2mm]

The discussion of principles of design makes it clear that there are essential differences between experiments and observational studies. In experiments, crucial aspects are under the investigators' control while in observational studies the features measured will largely have to be accepted as they happen to arise.  Cochran stresses however that, nevertheless, experiments and observational studies have much in common.\\[-2mm]

In particular, for the types of observational study he is discussing, the motivation is a search for causes.  Several variables may be viewed as treatments in a broad sense.  For instance, stronger positive effects may be expected for a set of new teaching methods, or stronger negative effects after exposure to higher levels of several risk factors for a given disease. When experiments are not feasible, the main aim is still to establish, as firmly as possible, the link with an underlying data generating process. Cochran states this as:  `A claim of proof of cause and effect must carry with it an explanation of the mechanism by which this effect is produced.' Thus, an underlying data generating process is to be scientifically explainable in the given subject matter context.\\[-2mm]

Some of the terminology has changed since the paper was written, but several key aspects remain essential for any planned study today:
\begin{itemize}
\item 	stating the main objectives of a study before the data are collected,
\item 	planning for well-defined comparisons and for one or several control groups,
\item 	thinking about the types of  measurements needed and how to assure their comparability,
\item 	specifying target populations and being aware of nonresponse as one reason for missing a target.
\end{itemize}

The relative importance of these aspects may differ in different fields of application. For example, in many areas of physics there is likely to be a secure base of background knowledge and theory, whereas in some types of social science research, this may not yet be the case. 
The broad approach to design must depend also on the time-scale and costs of a single investigation. Whenever new studies can be designed speedily and the data can be collected quickly and analyses are easily computed and interpreted, then a flexible approach with a sequence of simple studies may be feasible. But when the effort and time involved in any single study is considerable, all the above four points become essential for the study to become successful. A noteworthy example is the prospective study by Doll and Hill (1956) establishing cigarette smoking as a cause of lung cancer.\\[-2mm]

For experiments, R.A. Fisher (1926, 1935) had suggested, as principles of design, the need to avoid systematic distortions in treatment effects and the enhancement of the precision in estimates of effects.  He stressed also the value of considering several treatments simultaneously rather than one factor at a time. This gives the chance to see whether effects are substantially modified for particular levels of another factor or for level combinations of several factors,  that is to understand major interactions. More importantly, it may help to establish the stability of an effect under a range of conditions by showing the absence of major interactions. This idea carries directly over to observational studies.\\[-2mm]

However, to avoid systematic distortions, called often also `bias', is considerably harder in observational studies. In experiments, in addition to creating  laboratory-like conditions for obtaining  measurements for quantitative variables and observations for
categorical variables, the main tools are randomization, that is random allocation
of participants to treatment levels, stratification (called also subclassification or standardisation), the use of important covariates (in some contexts called concomitant variables) and blocking (which turns in observational studies into matching). \\[-2mm]

Clinical trials with randomized allocation of patients to treatments ideally may be regarded as experiments rather than observational studies. 
But in reality, distorted estimates of treatment effects can occur even in such clinical trials, for instance when relevant intermediate variables are overlooked, such as non-compliance of patients to some of the assigned treatments, or when there is a substantial undetected interactive effect of a treatment and a background variable on the
response, even though, by successful randomization, this background variable has
become independent of the treatment. Thus, Cochran's statement (on page 85) that  `in regard to the effect of x on y, matching and standardization remove all bias' cannot hold when one of the above mentioned sources of distortion for treatment effects is present.\\[-2mm]

When randomization is not an option, the next best approach is to design a prospective longitudinal study. But it may take a long time to see any results and these types of study are often expensive. They offer however the possibility of deriving and studying data generating processes. This option was not yet available in the 1970's except in the special situation of only linear relations and with responses that are affected one after the other, that is when path analysis, called recursive systems by Strotz and Wold (1960), is applicable. The importance of such an approach was rarely appreciated at that time; the textbook by Snedecor and Cochran (1966) was a notable exception.  \\[-2mm]

The direct generalisation of path analysis, to include other than linear relations and arbitrary types of variables, is to the directed acyclic graph (DAG) models. A more appropriate class of models for data generating processes are the recursive systems in single as well as joint responses, called traceable regressions; see Wermuth (2012), Wermuth and Cox (2013, 2014). In these models, several responses may be affected at the same time, such as for instance, systolic and diastolic blood pressure which are two aspects of a single phenomenon, namely the blood pressure wave. Both will for instance be influenced at the same time when patients receive a medication to reduce high blood pressure. \\[-2mm]

These sequences of regressions form one subclass of the so-called graphical chain models and they include DAG-models as a subclass. They often permit the use of a corresponding graph to trace pathways
of development and they may be compatible with causal interpretations. They also
take care of a main criticism of DAG-models regarding causal interpretations by
Lindley (2002): that DAG's do not include joint responses 
and therefore cannot capture many types of causal processes.
\\[-2mm]

In the last section of the present paper, there is a beautiful illustration of the suggestion `make your theories elaborate', given by R.A. Fisher when asked how to clarify the step from association to causation; see Cochran (1965). We fully agree that this step needs careful planning of studies and good judgement in interpreting statistical evidence. \\[-2mm]

In the meantime, some of our colleagues have derived a `causal calculus' for the challenging process of inferring causality; see Pearl (2014). In our view, it is unlikely that a virtual intervention on a probability distribution, as specified in this calculus, is an accurate representation of a proper intervention in a given real world situation. 
Their virtual intervention on a given distribution just introduces some conditional independence constraints and leaves all other aspects unchanged. This may sometimes happen, but experience from controlled clinical trials suggests that this is a relatively rare situation.\\[-2mm]

Even before the step to a causal interpretation, it is, as discussed below,  less clear that matching or some adjustment will always be beneficial in observational studies.  For instance, with pair-matched samples, no clear target population is defined, hence it remains often unclear to which situations the results could be generalized. Blocking in experiments and matching in observational studies clearly make the measurements in different treatment groups more comparable. And, it has been demonstrated explicitly, how with more homogeneous groups to compare, both sampling variability and sensitivity to other sources of distortions are reduced; see Rosenbaum (2005). \\[-2mm]

But for data looked at only after pair-matching, it becomes impossible to study dependences among the matching variables, in particular, to recognize an extremely strong dependence among them in a target population that could even lead to a reversal of the dependence of the response on this
treatment. In addition, if results for the dependence of a response are computed exclusively for explanatory variables other than the matching variables, then an
important interactive effect, of a treatment and a matching variable on the response, may get overlooked; for some examples see McKinlay (1977). \\[-2mm]

 The same holds for caliper matching, as defined in the present paper, and for a formal extension of it, called propensity-score matching by Rosenbaum and Rubin (1983). For a careful study and discussion of the large differences in estimated bias that can result with different choices of variables included in the propensity score and with different types of matching methods, see for instance Smith and Todd (2005). 
Similarly, any adjustment of estimates depends typically on how well the associated model is specified; see for instance Bushway et al (2007). For poor estimates or with some model misspecifications, adjustments may do harm instead of being beneficial. For approaches to move away from mere adjustments, see for instance Genb\"ack et al. (2014).\\[-2mm]

In all of these discussions of matching and adjustments in the literature, generating processes are rarely mentioned. But their importance was already stressed in the present paper even though at that time, more than 40 years ago, the corresponding  sequences of regressions, necessary for full discussions, had been studied intensively only for the very special situation of exclusively quantitative responses and linear dependences.\\[-2mm]

 Generating processes lead from background variables, such as intrinsic features of the individuals, via treatments and intermediate variables to the outcomes of main interest. In corresponding sequences of regressions, the dependence structure among directly and indirectly important explanatory variables is estimated and different pathways for dependences of the responses are displayed in corresponding regression graphs.  \\[-2mm]

Such graphs may be derived from underlying statistical analyses for a given set of data and they represent hypothetical processes that can be tested in future studies. In addition, consequences of any given regression graph can be derived. Consequences  that result after marginalizing over some of the variables or after  conditioning on other variables  in such a way that the conditional independences present in the generating process are preserved for the remaining variables, can be collected into a `summary graph' by using, for instance, subroutines in the program environment R; see Sadeghi and Marchetti (2012).\\

In this way, it will become evident which variables need to be conditioned on and such knowledge may possibly lead to a single measure for conditioning. Generating processes will point directly to situations in which seemingly replicated results in several groups, such as strong positive dependences, change substantially after marginalising over some of the groups, in some cases even turning positive into negative dependences. This can happen only when some of the grouping variables are strongly dependent. This well known phenomenon  has been named differently in different contexts, for instance as the presence of multicollinearity, as highly unbalanced groupings, or as the Yule-Simpson paradox.  Conditions for the absence of such situations have been named and studied as conditions for `transitivity of association signs' by Jiang et al. (2014).\\[-2mm]

With the dissemination of fully directed acyclic graph models, some more recent terminology has become common.  For instance, when an outcome has one important explanatory variable and there exists, in addition, an important common explanatory variable for both, the latter is a confounding variable and when unobserved, it is now  named an `unmeasured confounder' that may distort the true dependence substantially. Similarly, when an outcome has one important explanatory variable and another outcome depends strongly on both, then by conditioning on this common response, a distortion of the first dependence is introduced and is named `selection bias'.\\[-2mm]

In the current literature on `causal models', known to us, both these types of distortions are discussed separately.  A related phenomenon, for which a first example had been given by Robins and Wasserman (1997), is typically overlooked: by a combination of marginalizing over and conditioning on variables in a given generating process, a much stronger distortion, named now `indirect confounding', may be introduced than by an unmeasured confounder alone or by a selection bias alone. Parametric examples for exclusively linear dependences and graphical criteria for detecting indirect confounding, in general, are available.  The latter use summary graphs that are derived by marginalizing only; see Wermuth and Cox (2008, 2014).\\[-2mm]

The broad issues so clearly emphasized in the present paper remain central, challenging and relevant. That is to say, not only are firm statistical relations of particular kinds to be established, such as the estimation of treatment effects and of possibly underlying data-generating processes, but the statistical results need to be interpretable in terms of the underlying science. \\[-6mm]

\renewcommand\refname

\end{document}